# Temporal extrapolation of heart wall segmentation in cardiac magnetic resonance images via pixel tracking

**Arash Rabbani[1,2], Hao Gao[1], Dirk Husmeier[1]**
[1]School of Mathematics & Statistics, University of Glasgow, Glasgow G12 8QQ, United Kingdom
[2] School of Computing, University of Leeds, Leeds LS2 9BW, United Kingdom
a.rabbani@leeds.ac.uk; hao.gao@glasgow.ac.uk; dirk.husmeier@glasgow.ac.uk

***Abstract*** - In this study, we have tailored a pixel tracking method for temporal extrapolation of the ventricular segmentation masks in cardiac magnetic resonance images. The pixel tracking process starts from the end-diastolic frame of the heart cycle using the available manually segmented images to predict the end-systolic segmentation mask. The superpixels approach is used to divide the raw images into smaller cells and in each time frame, new labels are assigned to the image cells which leads to tracking the movement of the heart wall elements through different frames. The tracked masks at the end of systole are compared with the already available manually segmented masks and dice scores are found to be between 0.81 to 0.84. Considering the fact that the proposed method does not necessarily require a training dataset, it could be an attractive alternative approach to deep learning segmentation methods in scenarios where training data are limited.
***Keywords***: Pixel tracking, superpixels, ventricle, cardiac magnetic resonance images, image segmentation.

## 1. Introduction

With cardiovascular diseases responsible for around 30% of all global deaths [1], monitoring the patient's cardiac function is a crucial foundation for effective treatment. Cardiac magnetic resonance (CMR) imaging is a non-invasive imaging technique that can provide valuable information about the shape, size, movement, and scars of the patient's heart [2]. CMR cine images are a series of two-dimensional image frames that represent the movements of the heart in a full cardiac cycle from systole to diastole. In the clinics, manual segmentation of the patient's heart is regularly done by experts to calculate the ejection fraction and a few other important geometrical features. This process is time-consuming and prone to bias due to operator bias or random error [3]. Deep machine learning has already proved its capability for accurate segmentation of the heart wall in CMR images [4], however, this approach heavily depends on the quality of an extensive training dataset. In the present study, a pixel tracking approach is developed with few dataset requirements and an inherent capability for the extrapolation of segmentation masks in other time frames. Although motion tracking in CMR images has been widely used for the estimation of the myocardium strain [5, 6], ventricle mask tracking using a superpixels approach has not been practiced yet to the best of our knowledge. There are several image segmentation techniques to divide a grey-scale image into smaller cells that share common features, namely Watershed [7], Slic [8], and Superpixels [9]. Superpixels is a statistical method introduced in 2003 [9] that uses local image features such as contour, texture, brightness, and continuation to implement a random search for the best-unsupervised classification between the image regions. This process is done by maximizing the intra-region similarity and minimizing the inter-region similarity [9]. In this paper, we have utilized the superpixels segmentation technique to implement a pixel tracking approach.

## 2. Material and Method

In this study, we have used a dataset of CMR cine images publicly available under the name of Automated Cardiac Diagnosis Challenge (ACDC) [10]. The dataset has been originally published in 2017 and its training data contains CMR scans of 100 patients with 5 subgroups of normal, myocardial infarction, dilated cardiomyopathy, hypertrophic cardiomyopathy, and dilated right ventricle. In addition to raw short-axis images of the heart, this dataset contains manual segmentations of different elements within these images including the myocardium, left ventricle cavity, and right ventricle. For each patient, there are 20 to 30 sets of raw images available representing equally distanced time steps in a full cardiac cycle. However, manually segmented masks are only available at the end of diastole and end of systole. In this study, we have tailored a pixel tracking algorithm to extrapolate the manually segmented masks from the end of the diastole time frame across a full cardiac cycle. Also, to evaluate the performance of the pixel tracking, predicted masks at the end of the systole will be compared with the manual segmentations which is already available.

In this method, we initially extract the superpixels cells of the raw image. Then by overlaying the segmentation mask from the previous frame on top of the superpixels cells of the current frame, we decide that each of the cells belongs to which label. To assign a uniform label value to each of the current frame cells, the most repeated label value of the previous mask in each of the cells is calculated. This basically means that if the majority of the cell is occupied by a label, we assign that value to the whole cell. As far as spatial changes from one frame to another are not larger than the average size of the cells, the method is promising to work [11]. However, due to imaging noise and geometrical deformation of the objects, local noise and errors are inevitable, especially at the boundaries of an object. To deal with that random error two steps are suggested. First, to apply a gaussian smoothing filter on the raw image at each timeframe with a standard deviation of 0.5 to minimize grey-level noise. Second, median filtering on new masks at each time frame with the kernel size of 9 by 9 pixels. This process is summarized in Fig. 1 by assuming a synthetic noisy circular shape that gradually moves towards the right side of the canvas. The optimum values for the standard deviation and kernel size of the used filters are found by manual tuning. For this purpose, we have used the images from 25 patients in the ACDC dataset equally distributed between 5 subgroups. Also, another 25 patients are selected in the same manner to test the tuned method.

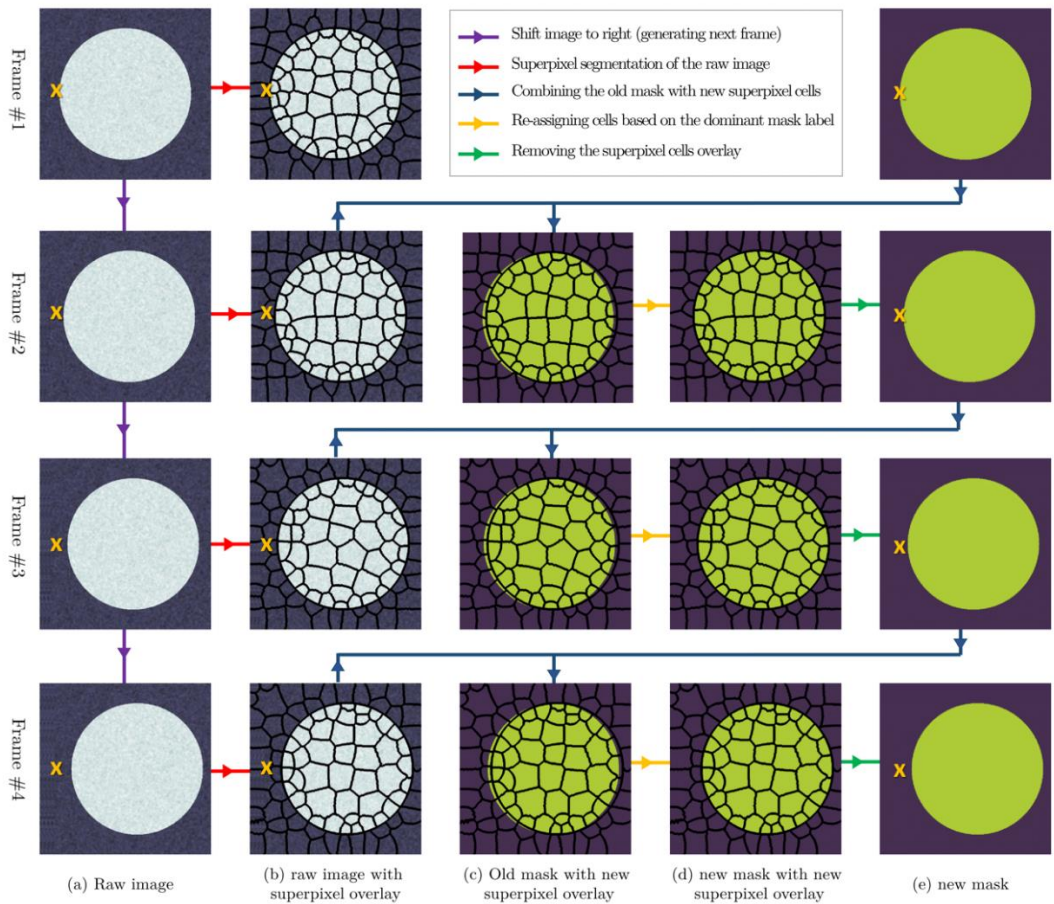

Fig. 1: A simplified example of the developed pixel tracking approach through 4 consecutive frames, the light grey circle is gradually moving towards the right side of the canvas and the orange cross is at a fixed location for reference.

Considering the fact that the heart moves in a periodic cycle, there are two pathways in the time domain that we can follow to reach from the end of diastole to the end of systole. The apparent choice is to track the heart ventricular movement in the same direction as time passes, and the second choice is to assume that the frames are happening in the reverse order. These two pathways lead to two different pixel tracking results for the end of systole as illustrated in Fig. 2. To improve the predictions, these two approximations can be combined to make an averaged mask and help to reach a more generalized result, which is possibly less dependent on the sequence of the tracked frames. The method we have used for averaging discretized masks is described in [12]. Sample results of the tracking and averaging processes are presented in Fig. 2. It can be seen that the predicted mask is visually similar to the ground truth, however, the similarity needs to be investigated quantitatively.

It is noteworthy that the presented methodology has been developed using the Image processing toolbox of the MATLAB programming language (R2021a) and the code and an example are publicly available at a GitHub repository for interested readers (https://github.com/ArashRabbani/VentricleTrack ).

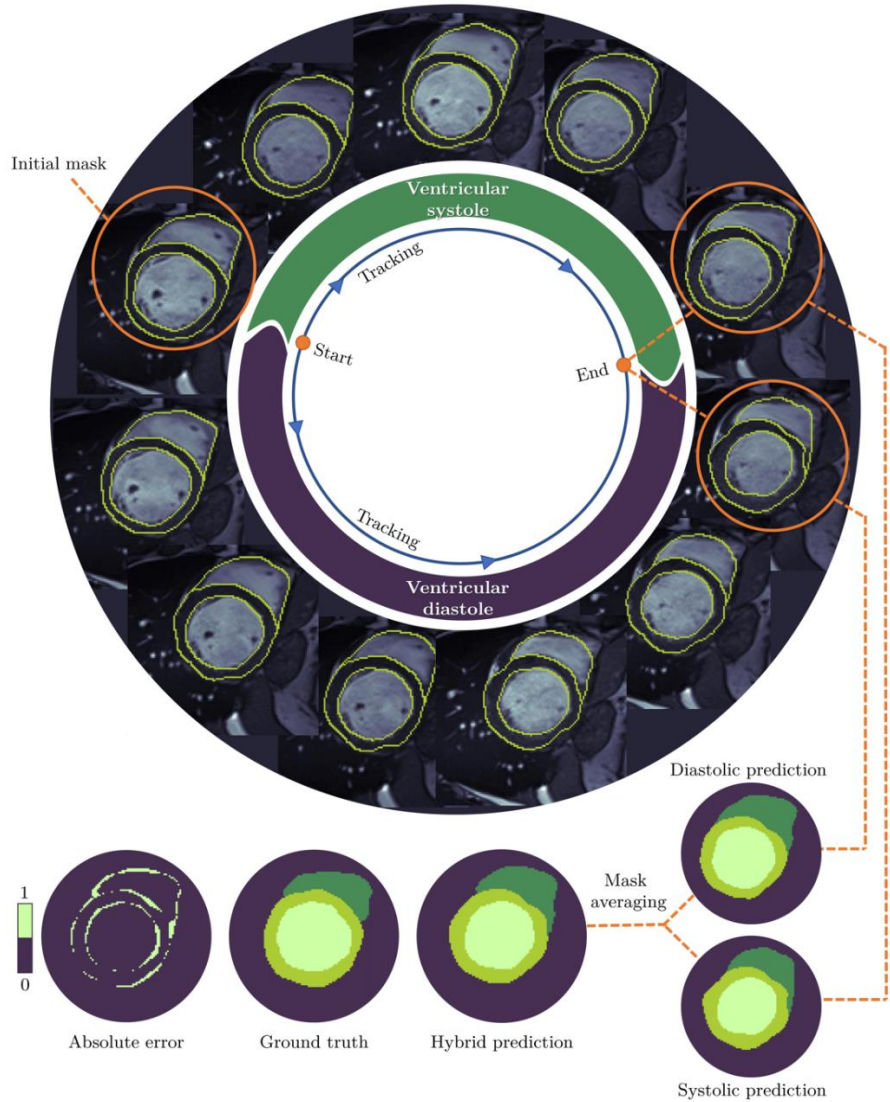

Fig. 2: Two temporal paths for pixel tracking from the end of diastole to the end of systole in the heart's pump cycle and combination of the two resulting masks for improving the stability of the prediction. The absolute error map shows the locations in which labels from the predicted and ground truth masks do not match.

## 3. Results and conclusions

As mentioned, we have used the available end-systolic segmentations to validate the proposed method of pixel tracking for temporal extrapolation of the ventricular image masks. This process has been done on the data from two groups of patients. First, the group of 25 patients which have been used for manual tuning of the method's hyper-parameters and, second, an out-of-the-sample group of 25 patients for testing the tuned method. In order to evaluate the accuracy of the predicted masks, we have used dice scores to compare them with ground truth images. The dice score has been traditionally used to evaluate the accuracy of image segmentation and it simply shows how well two image regions are covering each other when being superimposed [13]. It has a value between 0 and 1, and 1 means a perfect match. Fig. 3 illustrates the distribution of the dice score for 25 patients from each of the tuning and test groups. The scores are calculated for the segmentation of the right ventricle, myocardium, and left ventricle cavity in the range of 0.81 to 0.84. Although deep learning semantic segmentation has shown a higher accuracy [14] of between 0.9 to 0.99 for the same task, the fact that the presented method does not necessarily rely on a dataset to be trained on, makes it attractive for the cases with minimal availability of training data. In addition, considering the weakness of the machine learning models in extrapolation, the proposed method could be more competitive in the condition that the deep learning models have not been trained on any end-systolic images. Testing this hypothesis as well as developing a hybrid approach that combines both paradigms is an interesting topic for future research.

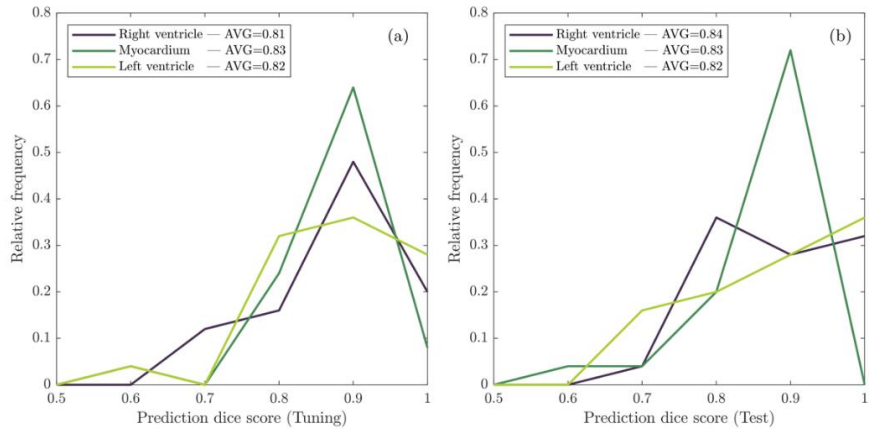

Fig. 3: Distribution of the prediction dice scores of tuning and test datasets for the right ventricle, myocardium, and left ventricle.

## Acknowledgment

This work was funded by EPSRC, grant reference number EP/T017899/1.